# Using the Built-in iPhone Body Tracking System for Neurological Tests: The Example of Assessing Arm Weakness in Stroke Patients: A Preliminary Evaluation of Accuracy and Performance

Vittorio Lippi[1,2][a], Isabelle Daniela Walz[2,3][b], Tobias Heimbach[2], Simone Meier[2], Jochen Brich[2][c], Christian Haverkamp[1][d] and Christoph Maurer[2][e]

[1]*Institute of Digitalization in Medicine, Faculty of Medicine and Medical Center, University of Freiburg, Freiburg im Breisgau, Germany*
[2]*Clinic of Neurology and Neurophysiology, Medical Centre, University of Freiburg, Faculty of Medicine, University of Freiburg, Breisacher Straße 64, 79106, Freiburg im Breisgau, Germany*
[3]*Department of Sport and Sport Science, University of Freiburg, Freiburg, Germany*
{vittorio.lippi, isabelle.walz}@uniklinik-freiburg.de, tobiasheimbach@gmx.net,
{simone.meier.neurol, jochen.brich, christian.haverkamp, christoph.maurer}@uniklinik-freiburg.de

Abstract: Timely treatment of stroke is critical to minimize brain damage. Therefore, efforts are being made to educate the public on detecting stroke symptoms, e.g., face, arms, and speech test (FAST). In this position paper, we propose to perform the arm weakness test using the integrated video tracking from an iPhone—some general tests to assess the tracking quality and discuss potential critical points. The test has been performed on 4 stroke patients. The result is compared with the report of the clinician. Although presenting some limitations, the system proved to be able to detect arm weakness as a symptom of stroke. We envisage that introducing a portable body tracking system in such clinical tests will provide advantages in terms of objectivity, repeatability, and the possibility to record and compare groups of patients.

## 1 INTRODUCTION

"Time is brain" – the later the treatment for a large vessel ischemic stroke, the more brain neurons are lost, and each hour costs around 3.6 years of normal aging (Saver, 2006). The magnitude of the insult plays a pivotal role in determining the course of action for rescue operations. It determines whether the patient is taken to a nearby hospital, typically for smaller infarctions, or a comprehensive stroke center (CSC), usually for larger infarctions (Václavík et al., 2018). Early thrombolytic therapy leads to a significantly better functional outcome in patients (Ospel et al., 2020). It is, therefore, crucial to assess the extent of the infarction as early as possible, but preferably when the emergency call is made.

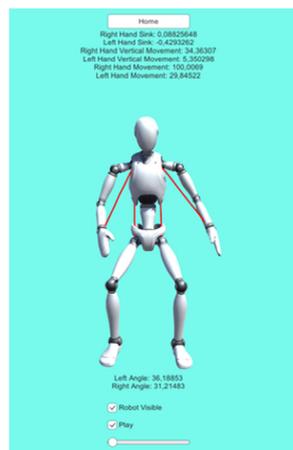

Figure 1: The "Robot" from the custom tracking application shows the kinematic as the system tracks it.

---

[a] https://orcid.org/0000-0001-5520-8974
[b] https://orcid.org/0000-0002-8033-1429
[c] https://orcid.org/0000-0001-6325-1892
[d] https://orcid.org/0000-0001-8165-4783
[e] https://orcid.org/0000-0001-9050-279X

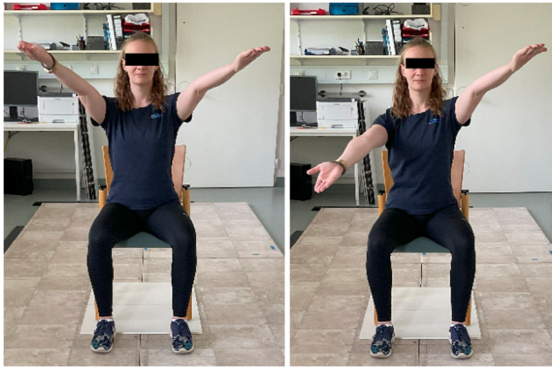

Figure 2: Healthy subject simulating a positive result for the test.

Efforts are therefore being made to educate patients, family members, and the general public (Saver et al., 2010). In rural areas, particularly when faced with a significant infarction, it becomes difficult to hit the within-3-hour window for an early thrombolytic therapy (Morris et al., 2000). Thus, efforts are underway to develop reliable methods for detecting substantial blockages, if possible, even from laypersons. In prehospital settings, the Face, Arms, Speech, Time (FAST) test is the most frequently employed scale to identify signs of stroke (Aroor et al., 2017; Budinčević et al., 2022). Especially the Arm Test in the FAST seems to be good for detecting a large vessel occlusion (LVO). Severe hemiparesis but also mono paresis are considered the most recognizable symptoms of an LVO stroke (Nakajima et al., 2004). The test is easy and straightforward to perform, does not require specific equipment, and can be performed everywhere, even on patients who cannot leave the bed. The test is qualitative by nature (as a result is "positive" or "negative") but is based on the judgment of the examiner. The introduction of an easy and quantitative measure (i.e., hand tracking) by only recording the arm test may provide advantages in terms of A faster detection from laypersons and easy documentation for emergency service. Nowadays, sensors in technologies have improved and are built into wearable devices like smartphones. Since 2020, the iPhone has been equipped with a LiDAR scanner capable of assessing three-dimensional scanning (Bhandarkar et al., 2021). So far, evaluation of the LiDAR sensor in mobile devices is still ongoing. The first results in scanning landscapes and little objects lead to the results that only larger objects can be measured (Teppati Losè et al., 2022). For objects >10 cm, an absolute accuracy of 1 cm is estimated. Therefore, we want to introduce a first easy and quantitative measure (i.e., hand tracking) by recording the arm test. This may provide advantages in terms of faster detection from laypersons and easy documentation for emergency services for people with stroke.

## 2 MATERIALS AND METHODS

### 2.1 The Smartphone Application

In order to perform the presented tests, an application has been developed with a unity engine, exploiting C code to interface with the ARKit library. The application can work on smartphones (iPhone) and tablets (iPad, Apple Inc., Cupertino, California). Specifically, the smartphone used in patient tests was the iPhone 12 Pro (2020, 128 GB, A14 Bionic chip, Model A2341), and the iPad (2021, 512 GB, 5$^{th}$ Generation, Apple M1 chip, Model A2378) was used for the preliminary tests. Both devices were equipped with LiDAR and True Depth capability. The preliminary test used the Captury system (§2.4) as a reference.

### 2.2 Tracking Library Overview

The Arkit Library for iPhone and iPad is designed for augmented reality. With this purpose, it provides the capturing of body motion in 3D: tracking a subject in the physical environment and visualizing their motion by applying the same body movements to a virtual character. The library relies on the camera and a LIDAR system, using lasers as a light source and the Time-of-Flight technique to measure distances (Gillihan, 2023). A skeleton (see the "robot" in Fig. 1) is fit on the tracked body, providing joint angles and link positions as output.

### 2.3 A Preliminary Tapping Test with the Captury System

In order to visualize the precision of the iPhone/Arkit system, a task with a defined hand movement has been used. Specifically, participants were told to touch two platforms with one hand as quickly as possible for 20 seconds. They were sitting on a chair while doing this arm-movement test. The platforms were positioned on the floor, aligned with the participant's feet. The platforms were 75 cm tall, and there was a 26 cm distance between them. This test was proposed by (Walz et al., accepted) and is based on the water-pouring task in the Fahn-Tolosa-Marin Clinical Rating Scale for Tremor (Fahn et al., 1988).

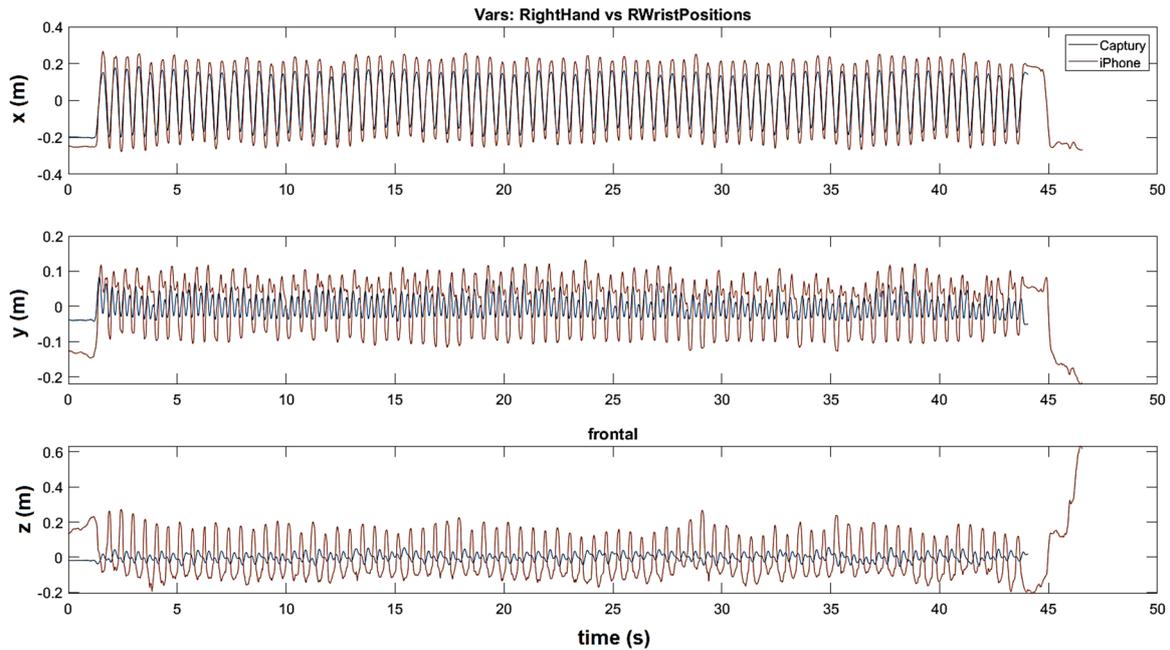

Figure 3: Tracking of the right hand during the tapping test. The plot shows one trial with the subject facing the camera. As the "skeletons" used in the *iPhone* and the *Captury* systems differ, the right hand tracked by the former is compared with the right wrist tracked by the latter. The components x,y, and z are expressed with respect to the camera and represent the left/right position (x), the vertical position (y), and the distance from the camera. The tracking is worse in the z-axis, as it is evident that there is a ripple effect. In this case (frontal view), the error is not much relevant in identifying the task.

We used the markerless system Captury (The Captury GmbH, Saarbrücken, Germany). to provide ground truth test movements (repetitive repeated arm-movements, as shown in Fig. 3). The Captury uses a method called visual hull and background subtraction to identify the shape of the subject. Then, it fits a skeleton on such a shape using an automatic scaling process. The camera system captures the movements at a rate of 50 Hz, giving us precise data on the positions of different body parts like the wrist, elbow, shoulder, hip, knee, and ankle and joint angles. In order to align the data between the two systems, the cross-correlation between the respective tracking of the moving hand is used:

$$f * g(\tau) = \int_{-\infty}^{+\infty} f(t)g(t+\tau)dt \qquad (1)$$

Specifically, the delay between the two systems is the $\tau$ that maximizes eq (1), where $f$ and $g$ are the profiles tracked with the two systems. The three-position components are used; hence, the product between the two functions is a scalar product $f^T g$.

All the computations were performed in Matlab (R2019b; MathWorks, Natick, Ma).

### 2.4 Participants

We recruited a total of 4 stroke patients for our study. All participants demonstrated their comprehension of the study procedures and provided written consent in accordance with the Declaration of Helsinki (World Medical Association, Declaration of Helsinki: Ethical principles for medical research involving human subjects, 2013).

### 2.5 The Arm Weakness Test

Before the main study, we conducted a pilot test, leading to minor changes in our test instructions. Specifically, patients were instructed to extend both arms in front of them, forming a V-shape to allow for better visibility of both arms. They were asked to fully extend their elbows and wrists while keeping their palms open and facing upward. Once patients closed their eyes, a 10-second recording was initiated to capture their movements. A clinical expert evaluated the patients' performance while simultaneously recording the motion data. The arm weakness was assessed via item 5 of the NIHSS rating scale (0 - no drift. 1 - drift, 2 - some effort against gravity, 4 - no movement). All recordings were made from a frontal perspective, with the patients seated upright.

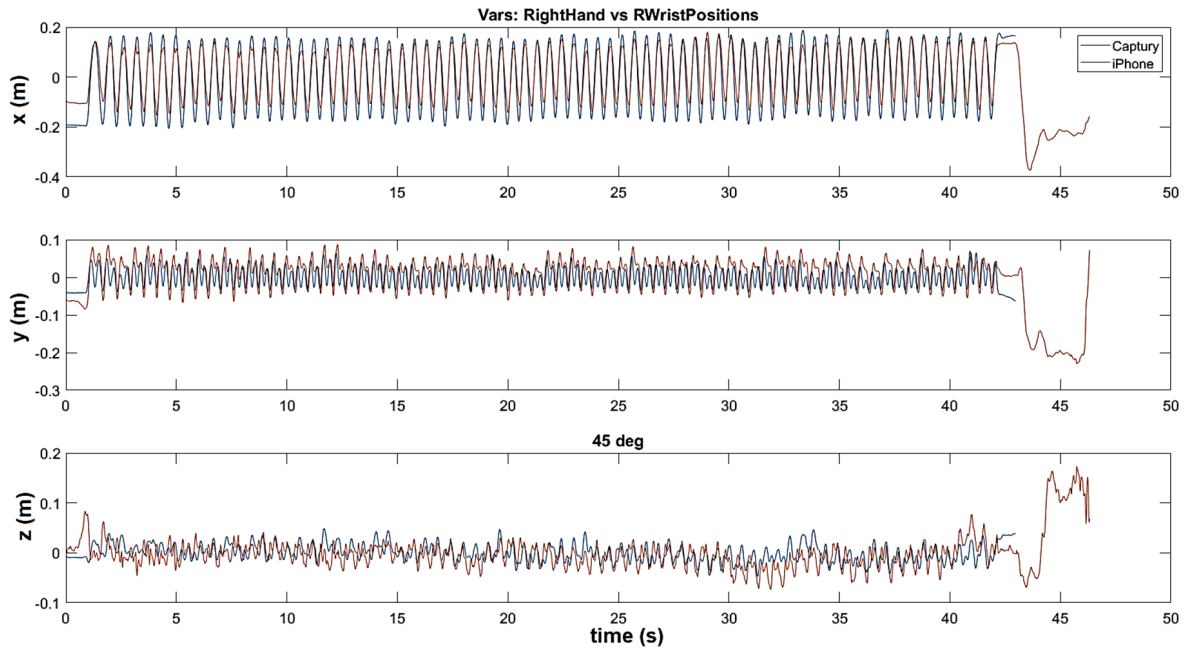

Figure 4: Tracking of the right hand during the tapping test. The plot shows one trial with the iPhone at 45° with respect to the subject. As the "skeletons" used in the *iPhone* and the *Captury* systems differ, the right hand tracked by the former is compared with the right wrist tracked by the latter. The components x,y, and z are expressed with respect to the iPhone camera and represent the left/right position (x), the vertical position (y), and the distance from the camera (z). The output of the Captury system has been rotated by 45° around the y-axis accordingly. The overall distortion is smaller than the one in Fig. 3, especially on the x-axis, as reported in Table 1.

## 3 RESULTS

### 3.1 Preliminary Test

The trials were performed with the iPhone in front of the subject and with an approximately 45° angle with respect to the subject. As the "skeletons" used in the iPhone and the Captury systems differ, the right hand tracked by the former is compared with the right wrist tracked by the latter. Fig. 3 shows the movement of the hand with the phone in front of the subject. In this configuration, the tracking of the right hand during the tapping task is better in the frontal plane than in the tracking of the depth (z-axis). The plot in Fig. 4 depicts a single trial where the iPhone is positioned at a 45° angle from the subject. In Fig 5, the same tracking with a 45° angle is shown for the left hand (not moving). The capture system demonstrates the absence of hand movement. However, the left hand, partially occluded, exhibits a rippling movement attributed to an artifact caused by the skeleton "shaking." Notably, the frequency of this artifact movement matches that of the intended movement performed by the subject. Overall, the iPhone tracking system identifies the movement of a body segment and its timing (i.e., the frequency is consistent with the one observed with the Captury) but has some error in movement amplitude in agreement with what was observed in early experiments (Reimer et al., 2021). Table 1 reports the difference between the Captury and the iPhone tracking in the first 30 seconds (after that, the subject stopped performing the movement, as is visible in Fig. 2,3, and 4) in terms of mean squared error. It is interesting to notice how, for the right hand, the 45° view produced a smaller tracking error, especially on the z-axis.

Table 1: Tracking error expressed as the standard deviation of the difference between the tracking performed by the iPhone and the Captury. The difference is computed over 30 seconds of tracking of the tapping test. The right and the left wrist are tracked. The right hand is performing the task; the right hand is kept in a relaxed position resting on the thigh.

|  | Frontal | | 45° degrees | |
|---|---|---|---|---|
|  | Right* | Left | Right* | Left |
| X | 0.0610 | 0.0499 | 0.0438 | 0.0571 |
| Y | 0.0640 | 0.0871 | 0.0292 | 0.0532 |
| Z | 0.1282 | 0.0711 | 0.0226 | 0.0510 |
| total | 0.1558 | 0.1230 | 0.0573 | 0.0932 |

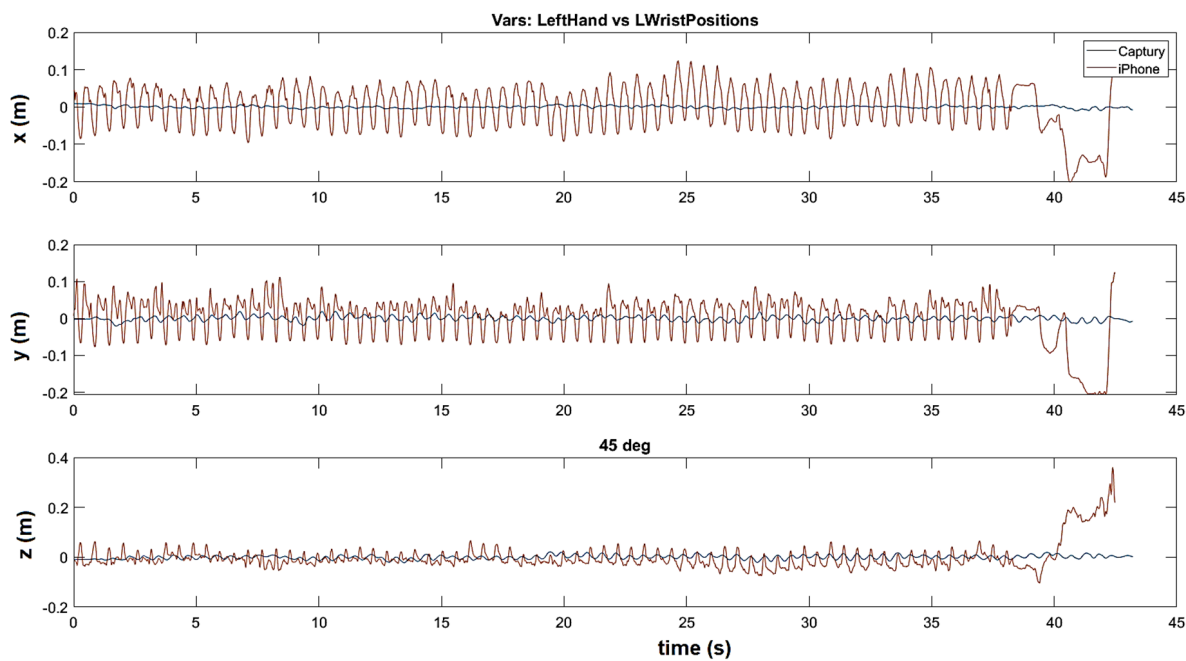

Figure 5: Tracking of the left hand during the tapping test. The plot shows one trial with the iPhone at 45° from the subject. The capture system shows that the hand is not moving. The partially occluded left hand is affected by a rippling movement that is an artifact due to the skeleton's "shacking." The frequency of the artifact movement is one of the voluntary movements performed with the right hand.

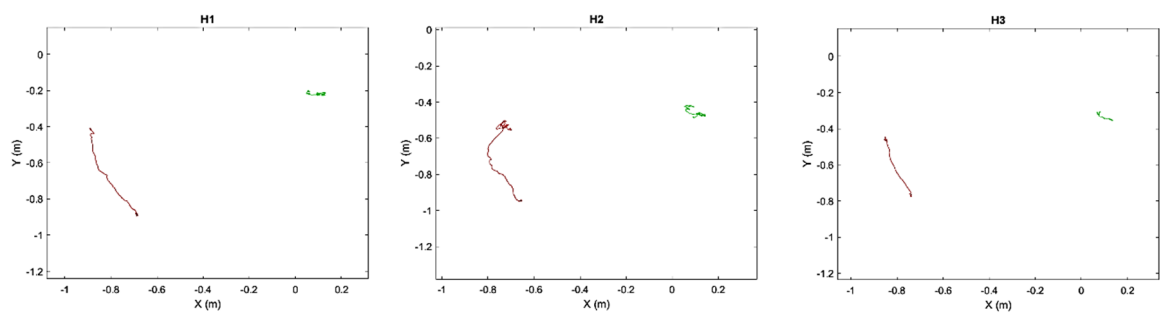

Figure 6: Three healthy subjects simulating a positive test. Hand trajectory tracking in the frontal plane from the iPhone.

## 3.2 The Arm Weakness Test

A preliminary check with healthy subjects simulating a test with positive results (Fig. 2) produced the trajectories shown in Fig. 6. It is evident how the range of motion of the hand that is dropping allows for positive identification of the movement and lateralization of the problem. The test with the four patients produced the outcome shown in Fig. 7 and described in Table 2. For the two subjects, S2 and S3, the iPhone tracking shows an evident drop (they were rated 1 by the doctor). The most severe case is represented by S4, rated 3. It was recorded in a lying position because of the impossibility of the patient to sit-stand upright. This produced less stable tracking and more "shaky" than the other cases. The excursion on the y-axis shows a large drift, although the patient returned to the original position. The tracking of hand pronation is not always reliable, as shown in Fig. 8. In some cases, e.g., subject 3 (S3), the pronation is clearly visible; in some recordings, it is ambiguous, like for subject 4 (S4). This is because, in general, the tracking of hand orientation is not always stable. Table 3 shows the amplitude of the vertical drop for the hands recorded in the tests (also the ones simulated by the healthy subject). It is evident that the hand affected by weakness is moving at a larger distance compared to the stable one. This allows for identifying the positive cases with a simple threshold, e.g., in the presented cases, a threshold of 30 cm would identify all the positive cases addressed by the doctor.

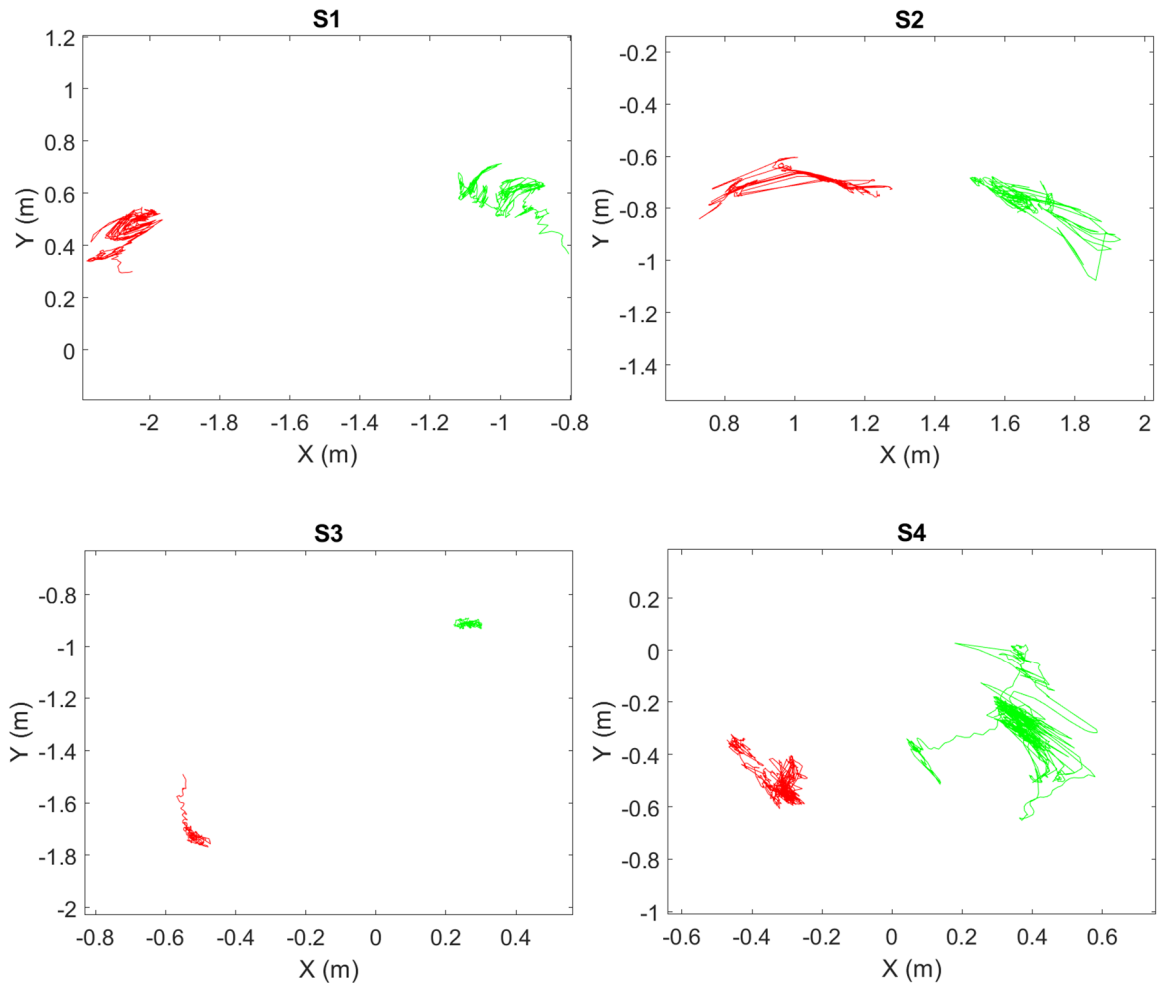

Figure 7: Test on stroke patients. In red, the right hand. In green, the left. The large vertical excursion (hand drop) is evident in the cases classified as positive by the doctor (i.e., S2, S3, S4).

Table 2: Report from the physician with the rating and some comments. Rating based on the NIHSS item 5 – Motor arm, 0 - no drift. 1 - drift, 2 - some effort against gravity, 4 - no movement).

| Patient ID | Rating | Comments |
|---|---|---|
| S1 | 0 | Pronation on the left, slightly bent elbow on the right, but then corrected it. |
| S2 | 1 | Pronation and minimal lateral descent of the right arm <10cm |
| S3 | 1 | Not sinking all the way to the bottom |
| S4 | 3 | |

## 4 DISCUSSION CONCLUSIONS AND FUTURE WORK

The results show that the iPhone body tracking system is suitable for identifying hand drops in patients and capturing movement features such as movement frequency for tasks like tapping in Fig 3,4, and 5.

Although equipped with a LIDAR that allows for direct measurement of depth, the iPhone can be prone to significant errors on the z-axis when the pose is ambiguous (Fig.3). This agrees with previous analysis reporting that, while the precision of the tracking is limited the system can provide useful features to be applied to patient examination (Reimer et al., 2021).

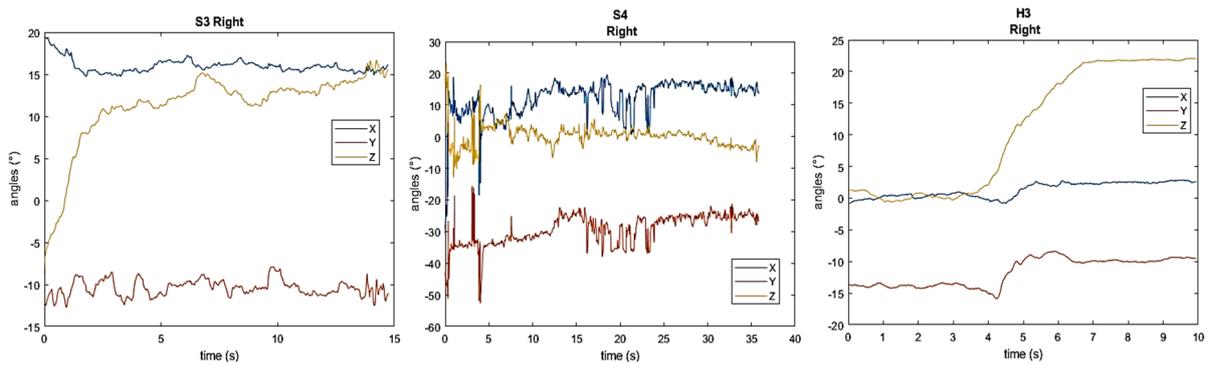

Figure 8: Tracking hand pronation. The figure shows the hand's orientation that is dropping during the test. On the left, the pronation is visible, accounting for a total rotation of about 20° around the z-axis. A similar outcome is visible on the right with the healthy subject H3. Subject S4 exhibits a very noisy trajectory from which it is impossible to assess the pronation properly.

Since the tasks examined in this paper require no feedback from the user, no considerations have been made about the delay that is not visible in the figures as the Captury and the iPhone recordings were aligned using eq. (1). For this purpose, it should be considered that the patient could perceive the delay from about 70 – 80 ms (Morice et al., 2008), and such an amount of delay can degrade the performance in the performed task (Lippi et al., 2010). This may be relevant in some applications as the iPhone tracking system is estimated to introduce around 120 ms of latency (Unlu & Xiao, 2021).

One intrinsic limitation of the iPhone is that being a single camera system, it is prone to the problem of occlusion (see the artifact movements of the hidden hand in Fig. 5). On the other hand, the experiments with different recording angles showed good flexibility in the possibilities to record the patient as required for the performed test. While the position tracking is good enough to identify the drop and be recorded as a quantitative measure, but the rotation was not always reliable in the examples. The integration of inertial units could improve overall precision, as shown in some applications, mainly using additional IMUs that the iPhone can read (e.g.:

Table 3: Vertical hand excursion was recorded with the healthy subjects and the simulated test.

| Subject | Left-hand drop (m) | Right-hand drop (m) |
|---|---|---|
| H1 | 0.0320 | 0.4934 |
| H2 | 0.0704 | 0.4478 |
| H3 | 0.0496 | 0.3314 |
| S1 | 0.2528 | 0.3479 |
| S2 | 0.2367 | 0.3982 |
| S3 | 0.0417 | 0.2777 |
| S4 | 0.7277 | 0.4672 |

in Kask & Kuusik, 2019; Monge & Postolache, 2018). To the best of our knowledge, there are no examples of test where the iPhone is handed to the patient for the test. The use of IMUs, on the other hand, would come at the costs of complicating the experimental test setup.

Overall, applying the iPhone as a diagnostic tool for neurological patients seems promising. One open point is storing the data from such tests in a way that is useful to make statistics on the patients and refine the model while preserving the subject's privacy. The design of an aggregated storage system will be the object of future work.